\documentstyle[11pt]{article}
\begin{document}
\title{Ferromagnetic ground states of the Hubbard \\
model on a complete graph}
\author{Mario Salerno \thanks{Email:SALERNO@csied.unisa.it}\\
\\
Department of Theoretical Physics \\
University of Salerno
\\ 84100 Salerno, Italy \\ and
\\ Istituto Nazionale di
Fisica della Materia  }
\maketitle
\begin{abstract}
We use group theory to derive  the exact analytical
expression of the ferromagnetic ground states of the
Hubbard model on a complete graph for arbitrary lattice sites
f and for arbitrary fillings $N$. We
find that for $t>0$ and for $N=f+1$
the ground state is maximally ferromagnetic with total spin
$S=(f-1)/2$. For $N > f+1$ the ground state is still ferromagnetic
but  becomes degenerate with respect to $S$.
\end{abstract}
\vskip 1cm
\newpage
An interesting aspect of the theory of itinerant
magnetism is the interplay between the kinetic energy
and the potential energy of the electrons as
a mechanism for the formation of ferromagnetic
ground states. Examples of such
mechanism leading to saturated ferromagnetism come
from the Hubbard model with one
hole in the strong coupling limit \cite{nagaoka66} and
from the Hubbard model with finite coupling but on
special graphs like the kagome lattice \cite{mielke92,tasaki92}.
The aim of the present paper is to show that
ferromagnetic ground states exist  also in the Hubbard
model on a complete graph (i.e. with unconstrained hopping)
for fillings higher than half filling and for arbitrary
couplings. The unconstrained hopping implies the invariance of the
Hamiltonian under the symmetric group $S_f$
(f being the number of lattice sites),
a feature which makes the model solvable for
clusters of arbitrary sizes \cite{ms1}.  In spite
of this simplification, the model
keeps all the complexity of the standard Hubbard model
coming from the Pauli exclusion principle and from
the $SU(2)$ invariance.
We hope that the knowledge of the exact ground states
of this model may be useful for understanding the local
properties of ferromagnetism in strongly correlated Fermi systems.
In a previous paper we have shown that for 
fillings $N$  below or equal to half filling ($N\leq f$), the
ground states are always non ferromagnetic
i.e. they are singlets or doublets depending
on $N$ being even or odd \cite{ms1}.
Here we extend this work by showing that
just above half filling ($N=f+1$ ) the ground state
is not degenerate with respect to the total spin $S$ and it
is maximally ferromagnetic with $S=(f-1)/2$.
For higher fillings ($N>f+1$), ferromagnetic ground
states exist but they are degenerate with respect to $S$.
This implies the existence of an hidden symmetry which
resemble the well known spin-charge symmetry of the 
standard Hubbard model \cite{woy83}. To analyze the system we
extend the  conjecture of ref. \cite{ms1} on the 
symmetry properties of the ground states 
under the action of $S_f$ and $SU(2)$, to the case of above half
filling. This conjecture was derived from computer analytical
diagonalizations of clusters of finite sizes by means of symbolic 
programs. 
Using this conjecture we derive analytical expressions for the ground
states energies  as a function of f and N and  we show
that the ferromagnetic ground states are always conducting.
Our analytical results are found in excellent agreement with
numerical diagonalizations this confirming the validity of our 
approach.

\noindent
The Hubbard model on a complete graph is written as
\begin{equation}
H=-t\sum_{\sigma, {i\neq j}}^fc_{i_\sigma
}^{\dagger }c_{j_\sigma }\;{+\;}
U\sum_i^fn_{i_{\uparrow }}n_{i_{\downarrow }},
\label{hub}
\end{equation}
where $c_{i_\sigma}^{\dagger },c_{j_\sigma }$ ,
$(\sigma =\uparrow$ or $\downarrow )$
are usual fermionic creation and annihilation operators.
In ref. \cite{ms1} we have shown how to use the
representation theory of the permutation group to construct
highest weight vectors of $SU(2)$ with definite $S_f$ symmetry
(for generalizations to arbitrary subgroups of $S_f$  
see ref.\cite{ms2}).
Here we just recall the main ideas leaving the  details 
to ref.\cite{ms1}.
Let us denote with the symbols $|3>, |2>,|1>,|0>$ the 
four states on a
given site (i.e. respectively, the doubly occupied, the single
occupied spin up and spin down states and the vacuum) and let
us introduce the quantum number
\begin{equation}
M=N + N_{\uparrow}.
\label{M}
\end{equation}
To construct the basis functions which span the irreducible
representations of the groups $S_f$ and $SU(2)$ we consider
all possible partitions
$(m_1,m_2,...,m_f)$ of $M$ into $f$ parts (compatible with
$N, N_{\uparrow}$), with $m_i=0,1,2,3$ and with
$m_1 \geq m_2 \geq ...\geq m_f$. Eigenmanifolds of $S^2, S_z$
and $S_f$ are then obtained  by filling the quanta
$m_i$ of  each  partition
in the boxes of the Young tableaux according to the
following rules.

\noindent i) The quanta must not increase
when moving from left to
right in a  row or when moving down in a column.

\noindent ii) The quanta referring to spin
up and spin down states ($m_i=1,2$) must not
appear more than once in a row.

\noindent iii) The quanta referring to doubly
occupied states or to empty states ($m_i=3,0$)
must not appear more than once
in a column.

\noindent
In order to select among the filled tableaux the
states  corresponding to highest weight vectors of
$SU(2)$ we use the following criterion.
The  application of the $SU(2)$ rising operator
$S^{+}$ to a state changes a spin down into a spin up
electron (1-2 flip). This implies
that the  states associated with tableaux satisfying the
filling rules also after 1-2 flips are necessary
states with $S>S_z$. The remaining tableaux will be automatically
annihilated by $S^{+}$ i.e. they will give
highest weight vectors of $SU(2)$.
By implementing the above rules on a computer
we have solved in exact form clusters of several sizes.
From these studies the following conjecture  
is derived for the case $t>0, U>0$.
\vskip .5cm
\noindent {\it{Conjecture}}:
\noindent For $N=f+1$ the ground state is maximally ferromagnetic
with $S=(f-1)/2$, it is non degenerate and it belongs to the
completely antisymmetric representation $\{1^f\}$.
For $N\geq f+2$, the ground state is L-fold degenerate with respect
to $S$ where $L$ is given by
\begin{equation}
L= {\frac{2f-N} 2} + \delta,
\label{L}
\end{equation}
with $\delta=1$ or $\delta={\frac{1}2}$ depending if N is even or odd.
The spins $S_j$ of the degenerate ground states are
\begin{equation}
S_j={\frac{2(f+1)-N} 2}-(j+1), \quad\quad  j=0,1,...,L-1,
\label{Sj}
\end{equation}
and the corresponding $S_f$ symmetries are given 
by tableaux $T_j$ of type
\begin{equation}
T_j \equiv \{ N-f\,,2^j\,,1^{2 (f-j)-N }\}, \quad\quad j=0,1,...,L-1.
\label{Tj}
\end{equation}
\vskip .2cm

\noindent
Using this conjecture we can derive analytical expressions  for
the ground states above half filling.
\noindent Let us first consider the case just above
half filling, $N=f+1$, where saturated ferromagnetism occurs.
According to our conjecture the ground state is a fully
antisymmetric state  with $S=(f-1)/2$. In order to be an
highest weight vector of $SU(2)$ it must have  $N_{\uparrow}=f$
and $N_{\downarrow}=1$ so that
from Eq. (\ref{M}) we have $M = 2 f+1$. There is only one way to fill
$2 f +1$ quanta into a tableaux of type $\{1^f\}$
according to our rules i.e. by putting
a double occupied state $3$ at the top  and $f-1$ spin up
states $2$ in the  remaining boxes as shown in Fig.1.
By applying the antisymmetrizer
operator to this tableau we obtain the maximally ferromagnetic ground
state as
\begin{equation}
|\psi>=\sum_{k=1}^{f} |k>,
\label{ferrstate}
\end{equation}
where $|k>$ denotes a state with $f-1$ spin up and one double
occupied state at site k. A direct calculation shows that $|\psi>$
is an  eigenstate of H with eigenvalue $-[t(N+2)-U]$
\begin{equation}
H |\psi> = -( t (N+2)-U )\, |\psi>.
\end{equation}
A similar analysis can be performed for higher  fillings.
We have that for $N>f+1$ the condition for highest  weight
vectors is
\begin{equation}
2 (f-j) - N = N_{\uparrow} - N_{\downarrow}, \quad j=0,1,...,L-1,
\end{equation}
which together with $N=N_{\uparrow} + N_{\downarrow}$
gives the following expression for the quanta in
Eq. (\ref{M}) to be filled in tableaux $T_j$:
\begin{equation}
M_j = N + f - j, \quad\quad j=0,1,...,L-1.
\label{Mj}
\end{equation}
It is remarkable that each tableau of type $T_j$
can be filled with $M_j$ quanta just in one manner.
We shall illustrate this with a concrete example. Let us take $f=5$
and $N=7$. In this case we have $L=2$ degenerate ground states
$|\psi_i>$, i=0,1,  of  spin $S_0 = 3/2$, $S_1 = 1/2$ and  given by
\begin{eqnarray}
\psi_0 &=&
|3\,3\,2\,2\,2> - |2\,3\,3\,2\,2> -
|2\,3\,2\,3\,2> - |2\,3\,2\,2\,3>,\\
\psi_1 &=& |3\,3\,2\,2\,1> - |1\,2\,2\,3\,3> +
|3\,3\,1\,2\,2> - |2\,2\,1\,3\,3> + \nonumber \\
& & |3\,2\,2\,3\,1> - |1\,3\,2\,2\,3> +
|1\,2\,3\,3\,2> - |2\,3\,3\,2\,1> + \nonumber \\
& & |2\,3\,1\,2\,3> - |3\,2\,1\,3\,2> +
|2\,2\,3\,3\,1> - |1\,3\,3\,2\,2> + \\
& & 2 \,(\, |3\,3\,2\,1\,2> - |2\,1\,2\,3\,3> +
|2\,3\,2\,1\,3> - |3\,1\,2\,3\,2> + \nonumber \\
& & |2\,1\,3\,3\,2> - |2\,3\,3\,1\,2> \,).\nonumber
\end{eqnarray}
\noindent
The corresponding filled tableaux are the ones reported in Fig. 2.
The energy of these states is readily obtained as
$E=<\psi_i| H|\psi_i>$, this giving $E=-3 t + 2 U$.
By extending this approach to arbitrary $f, N$
we find that the energy of the ground state are  given
(for $N>f$ and $t>0$) by
\begin{equation}
E(N,f)=-t (2 f - N) + U (N - f).
\label{egs}
\end{equation}
We have checked Eq. (\ref{egs}) by comparing with 
Lanczos numerical diagonalizations of clusters of several 
sizes up to f=16, always obtaining  a perfect agreement.
The fact that for $N>f+1$ the ferromagnetic
ground states  are degenerate with respect to
S implies the existence of an hidden  symmetry which
we can describe as follows.
The flipping  of a single occupied state from
spin up into spin down does not cost energy to the system if it is
accompanied by a spatial rearrangement of the electrons i.e. if the
electrons rearrange according to the next $T_j$ symmetry
given by Eq. (\ref{Tj}). In Fig. 3 we report the 
filled tableaux $T_j$ for fillings of type $N=f+k$ with $k$ even 
(for k odd one proceeds similarly). We note that $k$ is just the 
number of double occupied states and $0\leq j\leq L-1$ is 
the number of singly occupied spin down states present in 
the ground states.
The tableaux $T_0$ always gives the  maximally ferromagnetic 
ground state for a given filling. The others ground states 
are obtained from $T_0$ by removing spin up electrons 
from the first column and adding them as spin down electrons
in the second column. Note that this rearrangement of
the quanta is unique under the requirement of decreasing S 
while keeping the states highest weight vectors of $SU(2)$. 
From this we see that for $N=f+1$ the
ferromagnetic ground state cannot be degenerate 
($T_1$  cannot be an highest weight vector).
This accidental symmetry resemble the
spin-charge symmetry observed on bipartite lattices. To clarify this
let us recall that the spin-charge symmetry is related to the partial 
particle-hole transformation
\begin{eqnarray}
c_{i_{\uparrow }} &\rightarrow& c_{i_{\uparrow}}\\
c_{i_{\downarrow}} &\rightarrow& (-1)^i c_{i_{\downarrow}}^{+}
\label{pph}
\end{eqnarray}
which maps the $SU(2)$ algebra into the pseudo-spin $SU(2)$ 
algebra \cite{woy83}.
We note that the Hamiltonian (\ref{hub}) is not invariant
under this transformation  and the pseudo-spin does not commute
with $H$. In spite of this, 
we have that the above ground states  are highest weight vectors
of the pseudo-spin algebra so that they are highest weight vectors of
$SU(2) \times SU(2)$. To prove this statement we recall that the
generators of the pseudo-spin algebra are
\begin{eqnarray}
J^{+}&=&\sum_{i}(-1)^i c_{i_{\uparrow}}^{+} c_{i_{\downarrow}}^{+}\\
J^{-}&=&\sum_{i}(-1)^i c_{i_{\downarrow}} c_{i_{\uparrow}} \\
J_z&=& {\frac{N-f}2}.
\label{pspin}
\end{eqnarray}
From the filled tableaux in Fig.3 we see that the ground states have 
no empty sites so that they are all annihilated by $J^{+}$. 
Moreover, from 
\begin{equation}
J^2 = J^{-} J^{+} + J_{z}^2 + J_z  
\end{equation}
and from $ N=f+k $ we see 
that they are eigenstates of $J^2$ with $J = J_z= k/2$.
The fact that these ground states are invariant under pseudo-spin
rotations whilst the Hamiltonian is 
not, shows how extending far the 
consequence of the Pauli exclusion principle could be. 
\noindent
We also remark that  by a standard particle-hole transformation
the hamiltonian $H(t,U)$ goes into $H(-t,U) + (N-f)U$.
This implies that the above ground states for
$t,U>0$ and $N>f$, can be obtained  from the highest energy states
of $f-N$ holes simply  by changing the sign to 
t and shifting the energies
by $(N-f)U$. Finally note that the ferromagnetic ground states are 
all conducting since they satisfy the equation
\begin{equation}
E(N-1,f) - 2 E(N,f) + E(N+1,f)=0,
\end{equation}
as can be seen from Eq.\ref{egs}.
This result is in agreement with the thermodynamical analysis 
performed in ref. \cite{donvol}.

\vskip .5cm \noindent
\noindent{\bf {Acknowledgments}}

Financial support from the Istituto Nazionale di
Fisica della Materia, Unita' di Salerno, and from INTAS grant 
N. 93-1324 is acknowledged.

\newpage

\noindent
{\bf {Figure Captions}}
\vskip 1cm
\noindent Fig.1
Filled Young tableaux corresponding to the ferromagnetic
ground state of spin $S={{f-1}\over 2}$ at filling  $N=f+1$.
\vskip 1cm
\noindent Fig.2
Filled Young tableaux for the degenerate ground states of the case
$f=5$, $N=7$. The total spin $S$ of these states is respectively
(from left to right) $3/2,\,1/2$.
\vskip 1cm
\noindent Fig.3
Filled Young tableaux of type $T_j$ associated 
with degenerate ground
states at filling $N=f+k$ with $k$ 
even, $S_j=(f - 2j - k)/2$ 
and pseudo-spin $J=k/2$. Here $k$ denotes 
the number of double 
occupied states and $j$ the number of singly occupied 
spin down states  present in the ground states.

\newpage
\vskip 1cm

$$
\begin{picture}(200,300)
\put(100,250){\framebox(15,15){3}}
\put(100,235){\framebox(15,15){2}}
\put(100,160){\framebox(15,75){
\shortstack{.\\\\.\\\\.\\\\.}}}
\put(100,145){\framebox(15,15){2}}
\end{picture}
$$
\vskip .1cm
\centerline{Fig.1}

\newpage

\vskip 3cm

$$
\begin{picture}(200,300)
\put(0,250){\framebox(15,15){3}}
\put(15,250){\framebox(15,15){3}}
\put(0,235){\framebox(15,15){2}}
\put(0,220){\framebox(15,15){2}}
\put(0,205){\framebox(15,15){2}}
\put(100,250){\framebox(15,15){3}}
\put(115,250){\framebox(15,15){3}}
\put(100,235){\framebox(15,15){2}}
\put(115,235){\framebox(15,15){1}}
\put(100,220){\framebox(15,15){2}}
\end{picture}
$$

\vskip 1cm
\centerline{Fig.2}
\newpage
\vskip 3cm

$$
\begin{picture}(300,300)
\put(0,250){\framebox(15,15){3}}
\put(15,250){\framebox(15,15){3}}
\put(30,250){\framebox(20,15){.\,.\,.}}
\put(50,250){\framebox(15,15){3}}
\put(0,235){\framebox(15,15){2}}
\put(0,165){\framebox(15,70){
\shortstack{.\\\\.\\\\.\\\\.}}}
\put(0,150){\framebox(15,15){2}}
\put(0,130){${\bf{T}}_0$}
\put(100,250){\framebox(15,15){3}}
\put(115,250){\framebox(15,15){3}}
\put(130,250){\framebox(20,15){.\,.\,.}}
\put(150,250){\framebox(15,15){3}}
\put(100,235){\framebox(15,15){2}}
\put(115,235){\framebox(15,15){1}}
\put(100,180){\framebox(15,55){\shortstack{.\\\\.\\\\.}}}
\put(100,165){\framebox(15,15){2}}
\put(100,145){${\bf{T}}_1$}
\put(190,215){.\quad.\quad.}
\put(250,250){\framebox(15,15){3}}
\put(265,250){\framebox(15,15){3}}
\put(280,250){\framebox(20,15){.\,.\,.}}
\put(300,250){\framebox(15,15){3}}
\put(250,235){\framebox(15,15){2}}
\put(265,235){\framebox(15,15){1}}
\put(250,215){\framebox(15,20){
\shortstack{.\\.\\.}}}
\put(265,215){\framebox(15,20){
\shortstack{.\\.\\.}}}
\put(250,200){\framebox(15,15){2}}
\put(265,200){\framebox(15,15){1}}
\put(250,180){${\bf{T}}_{(f+k)\over 2}$}
\end{picture}
$$
\vskip .1cm
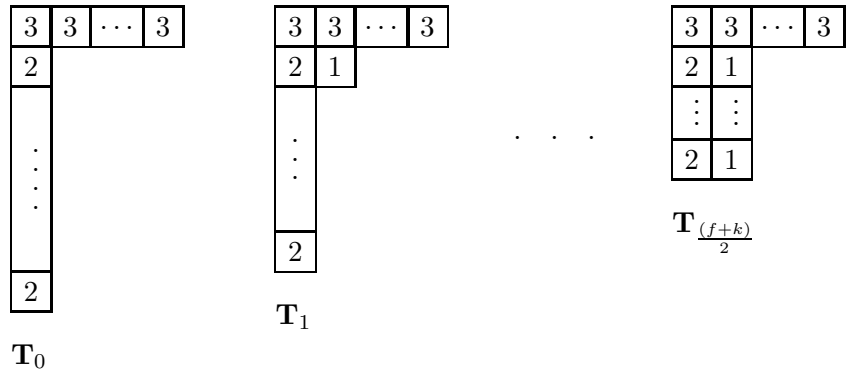
\centerline{Fig.3}

\end{document}